\documentclass[preprint,amsmath,amssymb]{revtex4-1}
\UseRawInputEncoding
\usepackage{amsthm,amssymb,amsbsy,amsmath,amsfonts,amssymb,amscd}
\usepackage[T1]{fontenc}
\usepackage{bm}
\usepackage{amsthm,amssymb,amsbsy,amsmath,amsfonts,amssymb,amscd}
\usepackage{graphicx}
\usepackage{bm}
\usepackage{epsfig} 
\usepackage{subfigure}
\usepackage{url}
\usepackage{comment}
\usepackage[dvipsnames]{xcolor}

\definecolor{darkblue}{rgb}{0.00,0.00,0.50}
\definecolor{darkgreen}{rgb}{0.00,0.50,0.00}
\definecolor{violet}{rgb}{0.4,0,0.3}
\definecolor{gris}{gray}{0.85}

\begin{document}

\title{Introduction to topological defects: from liquid crystals to particle physics}

\author{S\'ebastien Fumeron$^\ddag$}
\author{Bertrand Berche$^\ddag$}
\affiliation{$^\ddag$Laboratoire de Physique et Chimie Th\'eoriques, UMR Universit\'e de Lorraine - CNRS 7019, 54000  Nancy, France }


\begin{abstract}
Liquid crystals are assemblies of rod-like molecules which self-organize to form mesophases, in-between ordinary liquids and anisotropic crystals. At each point, the molecules collectively orient themselves along a privileged direction, which locally defines an orientational order. Sometimes, this order is broken and singularities appear in the form of topological defects. This tutorial article is dedicated to the geometry, topology and physics of these defects. We introduce the main models used to describe the nematic phase and discuss the isotropic-nematic phase transition. Then, we  present the different families of defects in nematics and examine some of their physical outcomes. Finally, we  show that topological defects are universal patterns of nature, appearing not only in soft matter, but also in biology, cosmology,  geology and even particle physics.
\end{abstract}

\maketitle

\section{Introduction}

Sometimes the deepest physics occurs just before our eyes, without really being noticed for decades except as a meaningless and minor accident of Nature. Open your hand and look closer at your fingerprints: most of the folds extend regularly, but sometimes, at the tip of our fingers or in the palm of the hand, they engage into strange circumvolutions serving no purpose, apart from fortunetellers and forensics. In fact, these seemingly innocent-looking patterns are some of the most fascinating objects in Nature: topological defects. These latter are at the crossroads of mathematics (topology, geometry) and physics (statistical physics but also cosmology, mechanical engineering...) \cite{poenaru1983elementary}.

\begin{figure}[h!]
\begin{center}
\includegraphics[height=6.9cm]{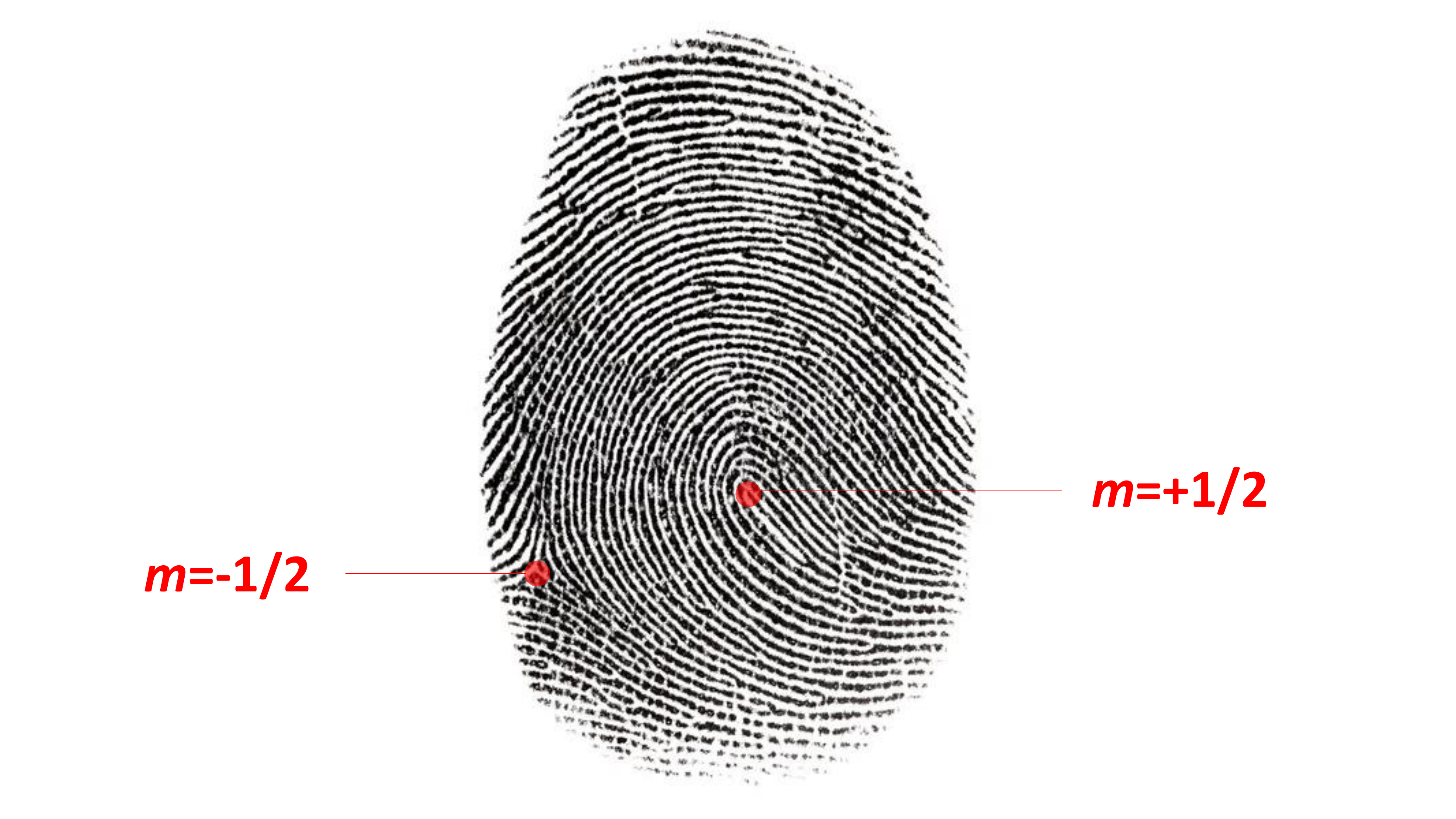} 
\caption{Patterns in human fingerprints. The values $m=\pm 1/2$ indicated there are explained later in this article.}
\end{center}
\end{figure}

The present tutorial is aimed as a self-contained introduction to topological defects in physics and it is split into three parts. The first section is designed as a primer on nematic liquid crystals. After some historical milestones, we review the main models that can be used to describe the nematic order (Frank -- Oseen, Maier -- Saupe, Landau -- de Gennes and Lebwohl -- Lasher). Then, we discuss the topology of the isotropic-nematic phase transition in connection with the formation of topological defects. 

In the second section, we present the basics of nematoelasticity and establish a classification of the different linear defects that can be observed in liquid crystals. The geometric description of defect lines in terms of Riemann manifolds is introduced and the general metric of a disclination is derived from a variational principle. Some outcomes on transport phenomena are discussed, in particular the possibility to generate Berry phases. 

The last section is a survey of different problems where the same linear defects occuring in liquid crystals have been recognized as the key factor. A particular emphasis will be put on biology, as tissues display an orientational order similar to the nematic order. We will see that defects play a pivotal role in morphogenesis and oncology. A particular attention will also be paid to cosmology, since singularities called cosmic strings seem to be in perfect correspondence  (formation, geometry, dynamics...) with the defect lines in nematics.

\section{A crash-course on liquid crystals}

\subsection{A story of carrots and sticks}

Liquid crystals have invaded our everyday life, from flat screens (liquid crystal displays or LCDs) to anti-counterfeiting technologies used in banknotes. As often in the history of science, their discovery started from quite a less utilitarian premise. In 1888, an austrian botanist, Friedrich Reinitzer, was trying to investigate the role played by cholesterol compounds contained in carrots. He successfully extracted crystals of cholesteryl benzoate and started to put them under a series of experimental tests. One of them consisted in measuring their melting point, which is expected to precisely and uniquely defined for pure crystals. Reinitzer observed a melting point at about 145.5$^{\circ}$ C for which cholesteryl benzoate crystals turn into a milky liquid... but against all odds, he found a second point at 178.5$^{\circ}$ C for which the turbid liquid becomes transparent. 

The perspective of a pure chemical compound having two melting points really bothered Reinitzer, who asked a renowned crystallographer, Otto Lehmann, about his observations. Lehmann was famous for inventing the ``crystallisation microscope", a apparatus coupling a regular optical microscope with two crossed optical polarizers and a thermal deck. He found that the turbid phase can flow like a liquid but was birefringent like an anisotropic crystal. He eventually launched an extensive program of experiments to discover other substances behaving like the turbid liquid. To describe such hybrid systems, he coined the terminology ``flowing crystals" (``fl\"ussige Krystalle") which later became ``liquid crystals".  

An important breakthrough was performed in 1907 by German chemist Daniel Vorl\"ander. Vorl\"ander realized that the optical and mechanical properties of the milky liquid originate from the elongated structure of the molecules composing it. Despite these advances, liquid crystals were not recognized as a noble research topic until the works of George Friedel, Fran\c ois Grandjean and Charles Mauguin, partly because of the vivid opposition of leading chemists such as Gustav Tammann, Walther Nernst, Georg Quincke and Tadeusz Rotarski, partly because of Lehmann's personality, a mix of priggishness and mysticism (mostly the result of his bounds with Ernst Haeckel). 

In 1922, Friedel made the most significant landmark in the development of liquid crystal. Being assemblies of microscopic sticks, liquid crystals display physical properties that prevent them from being either isotropic fluids or crystalline solids: they are new intermediate phases of matter, or mesophases. Friedel then classified mesophases into three main families, known as nematic, smectic and cholesteric phases. 

Nowadays, these classification has been enriched with many newcomers such as blue phases I, II and III, columnar phases, cubatic phases... It is now understood that these different mesophases are intimately related to the chemical structure of the mesogenic molecules composing them. Amphiphilic molecules have their head and tail displaying opposite chemical affinities. When dissolved in a solvent (typically water), the hydrophobic and the hydrophylic groups organize into membranes and micelles, which give rise to lyotropic mesophases (blue phases, cubic phases, hexagonal phases or lamellar phases) driven by the molecular concentration. Nematogenic molecules are partially rigid (generally because of phenyl groups) and bear electric dipoles (a carbonitrile group for 5CB, Schiff's base for MBBA...). They form temperature-driven (or thermotropic) mesophases (blue phases, columnar phases, smectic phases...) where the Van der Waals interactions are of the order of the thermal agitation. There also exist amphotropic mesophase, which share both lyotropic and thermotropic properties.

\subsection{Mathematical theories of the nematic phase}

For thermotropic nematics (on which we will focus hereafter), the molecules of liquid crystals may be described by long, neutral rigid rods, which interact through electrostatic dipolar or higher order multi-polar interactions. Maximization of entropy at high temperatures leads to a disordered phase in which all the molecule orientations are equally probable, independently of the directions taken by neighbouring molecules: this is the isotropic phase. At low temperatures a privileged orientation becomes favourable to minimize the molecular interactions, and various ordered structures may emerge spontaneously \cite{ANDRIENKO2018520,de1993physics}. The nematic phase is thus a compromise between the attractive van der Waals interactions that align rigid cores on average along the same direction (anisotropy) and the thermal agitation of the aliphatic chains increasing the mean steric hindrance (fluidity). 

When order occurs along one space dimension only, the system is said to be uniaxial. The preferential direction which emerges after averaging at the mesoscopic scales, defines a unit vector, $\vec n$, called the director field. As the dipole-dipole interactions tend to align molecules head-to-tail (dimeric head-tail structure was confirmed by X-ray diffraction in 5CB and 7CB \cite{leadbetter1975}), there is statistically no preferential arrangement of the molecules ends and $\vec n\equiv -\vec n$: this is called $Z_2$ symmetry. This dictates a natural order parameter in $\cos^2\theta$, with $\theta$ the angle between a molecule and the director field, rather than $\cos\theta$ in ordinary magnets (Zeeman interaction). Ideally, the order parameter should be normalized and mark a clear difference between a narrow distribution about $\pm \vec n$ and a random one. A choice fulfilling all these criteria is the scalar parameter proposed by Tsvetkov \cite{tsvetkov1942molecular}:
\begin{equation}
    S=\left\langle P_2(\cos\theta)\right\rangle=\frac{1}{2}\left(3\left\langle\cos^2\theta\right\rangle-1\right)
\end{equation}
where $P_2(x)$ is the second-order Legendre polynom\footnote{Note that the choice of the second Legendre polynomial here implicitly assumes 3-component $\vec n$ vectors. For 2-component vectors, a different choice is needed\cite{PhysRevE.72.031711}} and $\left\langle .\right\rangle$ is the ensemble average performed over all molecule in the nematic.  

The simplest way to describe the nematic phase is to rely on a mean-field theory, for which the microscopic details are blurred into a larger scale continuous field. The main continuum theories describing nematic liquid crystals are the Oseen -- Frank theory, the Maier -- Saupe theory and the Landau -- de Gennes theory. 

The Frank -- Oseen model is based on minimizing the free elastic energy: it is probably the simplest one to get the essential features of the nematic phase, even it fails at describing accurately the isotropic-nematic phase transition or the internal structure of defects (they have infinite energy). We will present it in more details in section \ref{topo-part2} to determine the outer structures of disclinations. 

In the Maier -- Saupe theory (1958), each molecule is in the mean field due both to long-range attractive pairwise interactions and steric repulsion with other molecules:
\begin{equation}
    V(\cos\theta)=-K\left\langle P_2(\cos\theta)\right\rangle P_2(\cos\theta) \label{potMS}
\end{equation}
where $K$ is the strength of molecular interaction, $\left\langle P_2(\cos\theta)\right\rangle$ is the mean field term and $P_2(\cos\theta)$ encompasses the angular dependance of the potential. In the canonical ensemble, the order parameter is thus given by an integral equation involving the orientational distribution function $f(\cos\theta)$  \cite{feng2005introduction}:
\begin{eqnarray}
\left\langle P_2(\cos\theta)\right\rangle&=&\int_0^1 P_2(\cos\theta)f(\cos\theta)d(\cos\theta) \\
f(\cos\theta)&=&\frac{\exp\left(-\beta V(\cos\theta)\right)}{Z}=\frac{\exp\left(-\beta V(\cos\theta)\right)}{\int_0^1 \exp\left(-\beta V(\cos\theta)\right)d(\cos\theta)}
\end{eqnarray}
with $\beta=1/k_B T$. Numerical resolution gives a phase diagram for $\left\langle P_2(\cos\theta)\right\rangle$ with respect to $T$ in good agreement with experimental datas \cite{feng2005introduction}: in particular, the transition temperature is found to be $T_c=0.00019 k_B T/K$  and the order parameter displays a small discontinuity at the transition.  

In mathematical models, one often considers the stronger approximation in which the molecules have a center of symmetry. Therefore, even though a molecule at space location ${\bf r}$ is idealized as a unit vector $\vec u({\bf r})$, an order parameter cannot be described by an ordinary vector like in ferromagnets and one has to consider the next invariant which is a second-rank tensor order parameter
\begin{equation}
    Q_{\alpha\beta}({\bf r})=\frac 1N\sum_{{\bf r}'}\bigl(u_\alpha({\bf r}') u_\beta({\bf r}')-{\textstyle\frac 13}\delta_{\alpha\beta}\bigr)
\end{equation}
The sum over the ${\bf r}'$'s extends inside a ball around $\bf r$, of size small, but large compared to microscopic scales. 
The $\vec u({\bf r})$'s live in a
three-dimensional space attached to each real space site and the $u_\alpha({\bf r})$'s are their Cartesian components. The ${\bf r}$'s which locate molecules are ordinary space vectors. Both ${\bf r}$'s and $\vec u$'s need not have the same dimensionality (e.g. a film of liquid crystal is naturally described by 2-component vectors  ${\bf r}$'s which describe the location of 3-component $\vec u$'s). 
The tensor $ Q_{\alpha\beta}({\bf r})$ has useful properties. It is symmetric and traceless, which reduces the number of its independent components from 9 to 5. It vanishes in the isotropic phase, and for prolate and oblate molecules, it respectively takes the forms
\begin{equation}
  [Q_{\alpha\beta}({\bf r})]=\begin{pmatrix}
  -1/3 & 0 & 0 \\
  0 & -1/3 & 0 \\
  0 & 0 & 2/3
  \end{pmatrix}
  \quad\hbox{and}\quad 
  [Q_{\alpha\beta}({\bf r})]=\begin{pmatrix}
  1/6 & 0 & 0 \\
  0 & 1/6 & 0 \\
  0 & 0 & -1/3
  \end{pmatrix}
\end{equation}
in the principal axes basis.

Being a local order parameter, the object $Q_{\alpha\beta}({\bf r})$ allows for a mean-field description in terms of an expansion of the free energy density known as Landau -- de Gennes theory. The approach is based on the assumption that in the vicinity of the phase transition where the order parameter vanishes, it is a small parameter and the free energy density can thus be expanded in powers of the order parameter. Since there are three independent invariants made from the tensor $Q_{\alpha\beta}$ (these are $\hbox{Tr}\ \!(Q)$, $\hbox{Tr}\ \!(Q^2)$ and $\hbox{det}\ \!(Q)$), the expansion is built from three leading terms,
\begin{equation}
    F=F_0+\int d^3r\Bigl( 
    {\textstyle \frac 12}A(T)Q_{\alpha\beta}({\bf r})Q_{\beta\alpha}({\bf r})
    +{\textstyle \frac 13}B(T)Q_{\alpha\beta}({\bf r})Q_{\beta\gamma}({\bf r})Q_{\gamma\alpha}({\bf r})
    +{\textstyle \frac 14}C(T)(Q_{\alpha\beta}({\bf r})Q_{\beta\alpha}({\bf r}))^2
    \Bigr).
\end{equation}
The first coefficient changes its sign at a temperature called $T^*$, $A(T)=A_0(T-T^*)$, and $B$ and $C$ are essentially independent of the temperature and their values can be approximated by those taken at $T^*$. The small elastic constant limit of Landau -- de Gennes theory converges to Frank -- Oseen theory \cite{majumdar2010landau}. Additional terms are not required if the coefficient of the higher order term, $C>0$, which ensures stability of the ordered phase. Further analysis shows that a first order transition takes place at a temperature $T_c$ slightly above $T^*$. This phenomenological theory is thus consistent with experimental evidence of a weak first-order transition (in three dimensional systems).

In real systems, the order parameter is slowly varying in space and additional terms in the expansion are required, which involve gradients of the order parameter. This is usually referred to as Ginzburg -- Landau theory then. This is particularly important in the presence of topological defects which impose specific spatial variations of the molecules orientations in the ordered phase.

The description of the nematic phase has also been extended to non mean-field approaches. In this perspective let us mention lattice models studies.
As discussed above for the order parameter, in the nematic phase, one can measure the deviation of molecules individual orientations
with respect to the director by the scalar product \hbox{$\vec u_i\cdot\vec n=\cos\theta_i$} -- where the $i$'s are now lattice sites (they play the role of the ${\bf r}$'s previously) -- but
due to the additional local $Z_2$ symmetry which identifies ``heads'' and ``tails'' one cannot distinguish between opposite directions. As a consequence, $\cos(\theta_i)$ vanishes on average while $\cos(\theta_i)^2$ does not and is thus more appropriate for a possible local interaction term. In the disordered phase, the angles are furthermore measured with respect to any arbitrary direction, and the thermal average leads to $\cos\theta_i=0$ and
$\cos(\theta_i)^2 = 1/3$. The quantity $\cos(\theta_i)^2 - 1/3$
thus represents a convenient scalar nearest neighbor interaction energy. 
In the literature on liquid crystals, one usually defines this by the second Legendre polynomial,
$\phi_i=P_2(\cos\theta_i)$, like for the order parameter.
This definition suggests to consider lattice Hamiltonians to describe the nematic transition, e.g. the so-called Lebwohl -- Lasher model
\begin{equation}
    -\frac H{k_BT}=\frac J{k_BT}\sum_{(i,j)} P_2(\vec u_i\cdot\vec u_j)
\end{equation}
where the sum is usually restricted to nearest-neighbours in the spirit of studies of critical phenomena. The constant $J$ is an empirical interaction parameter. 
This model is also called $RP^2$ model, and except for the $\cos(\theta_i)^2$, it looks very similar to the Heisenberg model. Nevertheless, the latter model exhibits a second-order transition in three dimensions while the Lebwohl -- Lasher model displays a first-order one.

Let us stress that the situation sketched above does not apply directly in two-dimensional liquid crystals. The question of the nature of the transition of the Lebwohl -- Lasher is still debated nowadays~\cite{FARINASSANCHEZ2003461,CMP2009,arxiv.2202.07597}, but the role of topological defects there is not under discussion. The situation is either similar to the Kosterlitz -- Thouless phase transition~\cite{2002dgcm.book.....N,berche2002correlations} which is governed by unbinding of topological defects pairs,
or to that of the Heisenberg model for which there is no transition at all, due to the instability of these defects. In all cases, two or three dimensional systems, the presence of topological defects is of paramount importance. 

\subsection{The isotropic-nematic phase transition for pedestrians}

To grasp the essential features of the isotropic-nematic phase transition, we work with the Maier -- Saupe model and compare the thermal agitation of the aliphatic chains (increasing the mean steric hindrance, e.g. fluidity) to the interaction potential (\ref{potMS})  that align rigid cores on average along the director field (anisotropy). Maier -- Saupe's leads to similar results than Landau -- de Gennes theory, but from rather simpler premises. The nematics in consideration hereafter are thermotropic, e.g. the external parameter driving the transition is temperature \footnote{Other possibilities include lyotropic nematics: the external parameter is the concentration of nematogens in a solvent and the most adequate description is Onsager's model.}.

\begin{figure}
\begin{center}
\includegraphics[height=8.9cm]{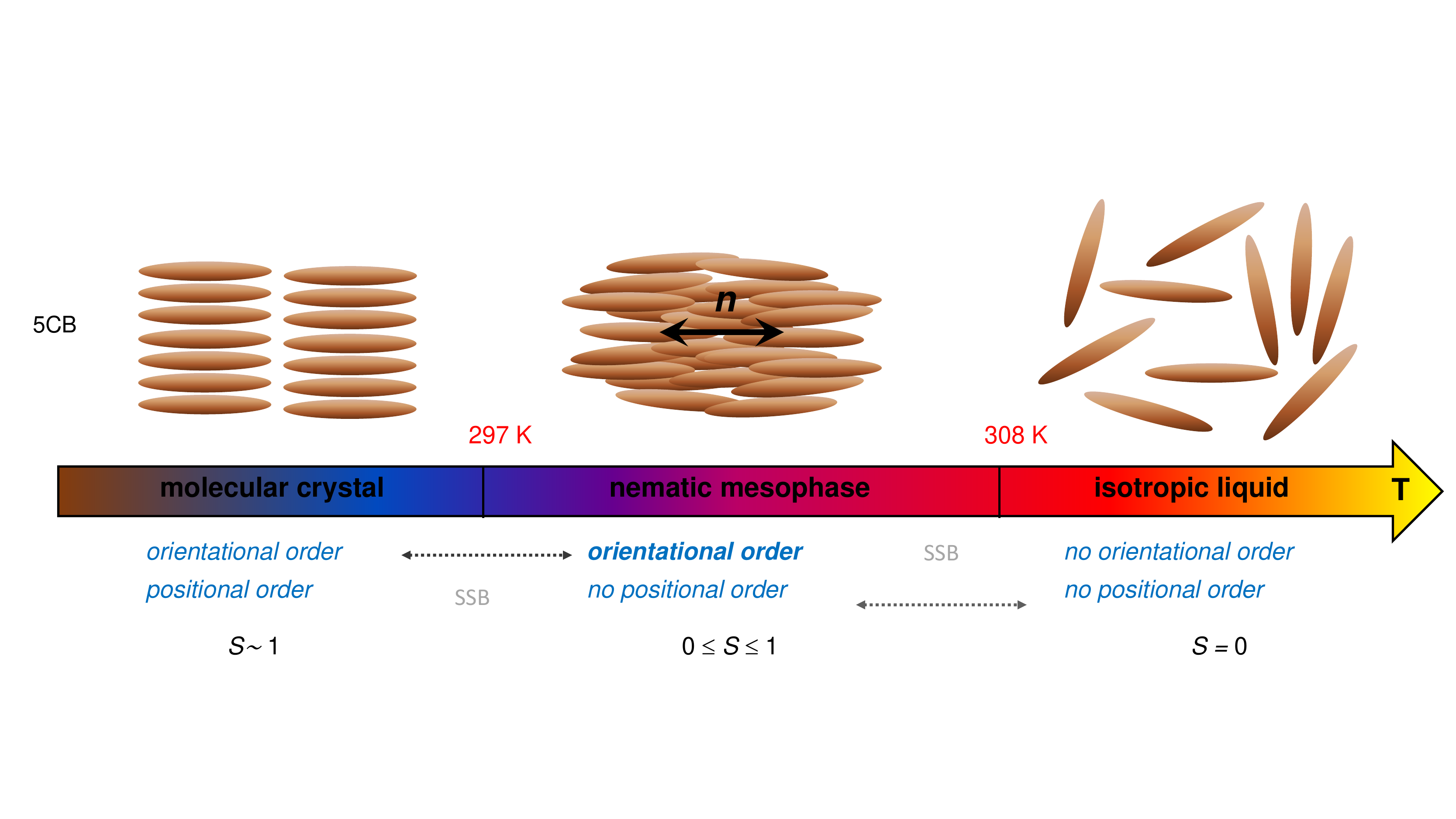} 
\caption{Phase transitions with 5CB.}\label{INPT}
\end{center}
\end{figure}

At low temperatures, thermal agitation is weak: ough to the low steric hindrance, nematogens are closer and van der Waals interactions prevail: the phase is that of a molecular solid (with sometimes a smectic phase in-between). The corresponding symmetry groups are discrete. At high temperatures, thermal agitation prevails over molecular interactions: the nematogens are distant from each other and they form an isotropic fluid. The rotational symmetry group of the phase is $SO(3)$, the group of rotation in three dimensions. Within the intermediate range of temperatures, with a symmetry axis given by $\vec{n}$: its symmetry group is therefore $SO(2)\times Z_2=O(2)$ (Fig \ref{INPT}). Therefore, the transition from isotropic to nematic phase involves a spontaneous symmetry breaking.

A thorough inspection of this transition reveals that it is a 3-step nucleation mechanism involving 1) the formation of small spherulites where an orientational order arises without correlation with each other, 2) the growth of the ordered domains and 3) the formation of threads when spherulites have mingled (see \cite{kleman2006topological} for self-speaking pictures of this process). We also saw that Maier -- Saupe theory and the Lebwohl -- Lasher model predict a weak discontinuity of the order parameter at $T_c$.  Measurements on different nematic samples also showed the existence of low latent heats, which suggests a first-order transition, but as pretransitional effects (decay of the dielectric constant with $T$ close to $T_c$) to have also been reported \cite{van2005weakly}, the isotropic-nematic phase transition is in fact weakly first order. Yet having cleared up the nature of the transition does not explain the presence of the threads. This is the object of the next section.

\subsection{Topology of the isotropic-nematic transition}

Topology is the branch of mathematics focusing on the properties that remain unchanged when a topological space is ``smoothly deformed" (neither torn apart nor punctured). When a topological property changes, it occurs by integer steps, not gradually. Algebraic topology (Poincar\'e' s former \textit{analysis situs}) seeks algebraic invariants (numbers, abelians groups, rings...) to study and classify topological spaces into equivalence classes. These objects may characterize properties such as the connectedness, the number of holes, the existence of boundaries... Two manifolds are topologically equivalent or homeomorphic if there exists a bijective and continuous map between them. These two manifolds have to be of the same dimension, as required by Brouwer's invariance of domain theorem. Intuitively, it corresponds to a continuous deformation with no gluing or tearing. Putting it colloquially, for an algebraic topologist, a doughnut is the same thing as a teacup  (an object with one hole), but a doughnut is different from a french pretzel (3 holes). 

Homotopy provides a weaker notion of equivalence between topological spaces than homeomorphism. It corresponds to a continuous deformation where bijectivity is not preserved, i.e. gluing, shrinking or fattening the space is allowed: for instance, in $\mathbb{R}^2$ a loop (dimension 1) is homotopic to a point (dimension 0). The first homotopy group (or Poincar\'e group, or fundamental group) is denoted as $\pi_1(\mathcal{M})$ and it tests if all closed curves (loops) on a manifold $\mathcal{M}$ are homotopic to a point. If it is the case, the fundamental group is trivial $\pi_1(\mathcal{M})=I$. When does this fail? When there are holes ! In the aforementioned example, $\pi_1(\mathbb{R}^2)=I$ but $\pi_1(\mathbb{R}^*{}^2)\neq I$: removing the origin creates a 0D-hole in the 2D-manifold. Technically, we just used the first homotopy group to test the simply-connectedness of $\mathbb{R}^*{}^2$. When $\pi_1(\mathcal{M})\neq I$, there are equivalence classes of homotopic loops sharing the same winding number (Whitney -- Graustein theorem). The winding number of a regular curve is the number of times the tangent vector fully rotates counterclockwise when going once around the curve. For $\mathcal{M}=\mathbb{R}^*{}^2$, $\pi_1(\mathcal{M})=\mathbb{Z}$: the equivalence classes consist in loops turning \textit{n} times around the origin (clockwise for $n<0$, counterclockwise for $n>0$). 

\begin{figure}
\begin{center}
\includegraphics[height=8.9cm]{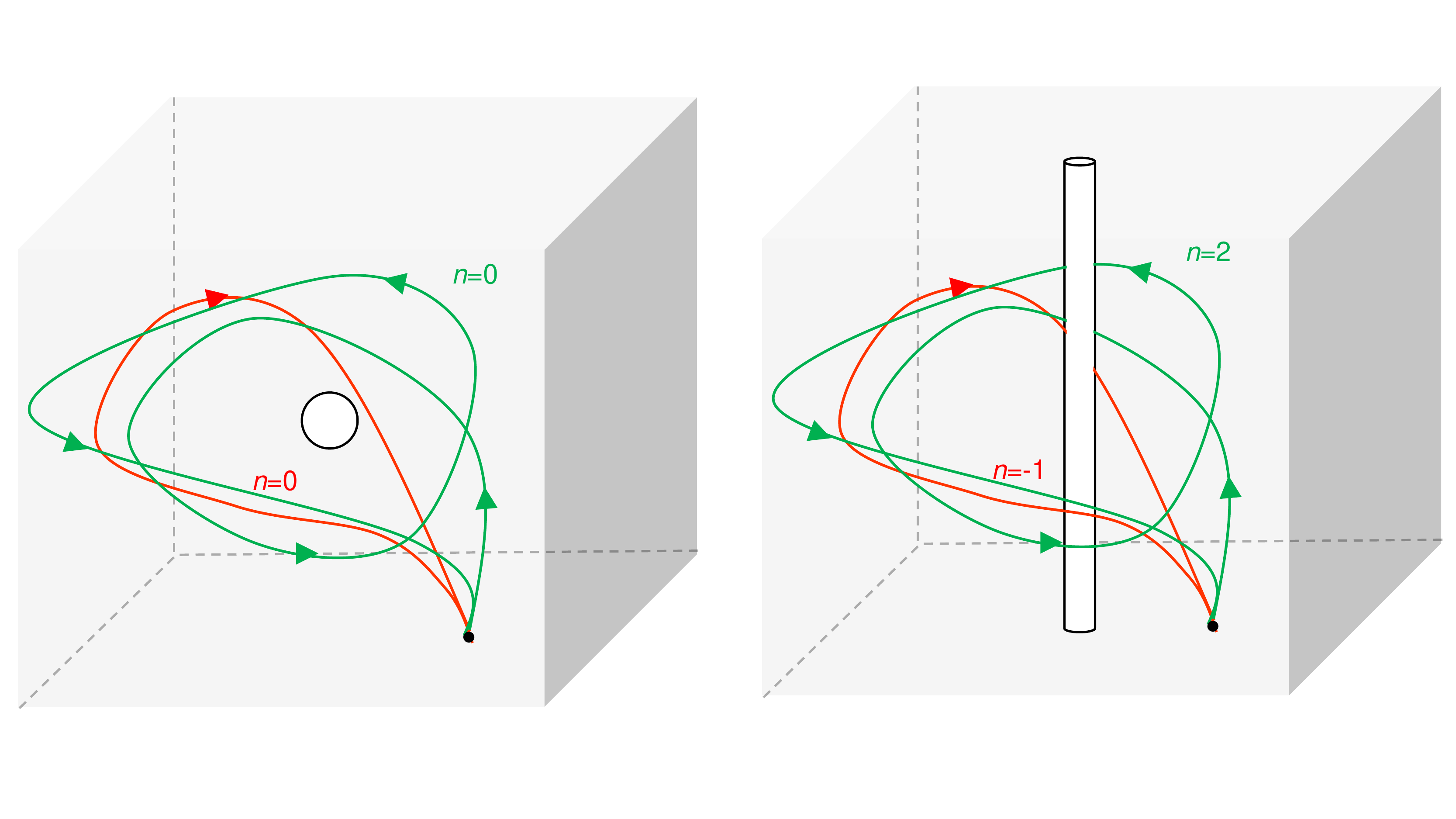} 
\caption{Left: Trivial closed loops around a 0D-hole (white region, spatial extent is for convenience). Right: Some equivalence classes around a 1D-hole (white region, spatial extent is for convenience.)}\label{lasso}
\end{center}
\end{figure}

They are many other topological properties than can be tested from homotopy groups: for instance, $\pi_0$  tests if the topological space is (pathwise) connected, i.e. if for any pair of points, one can find a path between them that remains in the topological space. Intuitively, it corresponds to the notion of a space that is in one whole piece. In that respect, $\pi_0(\mathbb{R}^2{}^*)=I$ but $\pi_0(\mathbb{R}^2-\{x+y=0\})\neq I$: thus, $\pi_0$ tests the existence of 1D-hole in the 2D-manifold. It has to be emphasized that the content of the homotopy group is strongly dependent on the dimension of the manifold: indeed, a 0D-hole (or point defect) in 3D-manifold can never be lassoed (see \ref{lasso}) and its fundamental group homotopy is trivial. For $\pi_1(\mathcal{M})$ to be non-trivial, the hole has to be 1D, e.g. a line defect (see Fig. \ref{lasso}). As a rule, we will bear in mind that in dimension \textit{p}, if the $k^{th}$ homotopy group $\pi_k(\mathcal{M})$ is non-trivial, then topological defects of dimension $p-1-k$ appear. 

Now, what is the connection of homotopy with phase transitions? This idea is simple: homotopy can predict the kind of defects that can appear after the phase transition from its symmetry-breaking pattern. More precisely, during a phase transition with a symmetry breaking pattern $G \longrightarrow H$, defects arise according to the topology of the order parameter space defined as the coset $\mathcal{M}= G/H$. Applications of homotopy to condensed matter have been investigated in many papers \cite{volovik1976line,kleman1977classification,michel1978topological,michel1980,volovik1983topological,kurik1988,kleman1989defects}. For the isotropic-nematic phase, $\mathcal{M}= SO(3)/O(2)=S^2/Z_2$: the order parameter space consists in a 2D-sphere having its antipodal points identified. Such object is known as the real projective plane and can be visualized from its immersion in 3D space, the Boy surface. The contents of the different homotopy groups are (\cite{michel1980}):
\begin{itemize}
    \item $\pi_0(\mathcal{M})=I$: one cannot observe surface defects in nematics.
    \item $\pi_1(\mathcal{M})=Z_2$: one can observe line defects in nematics, which are precisely the thread-like structures we were trying to explain since the previous section. These objects are generically called disclinations and Friedel give its name to the nematic phase after them (in greek $\nu\eta\mu\alpha$ means thread).
    \item $\pi_2(\mathcal{M})=\mathbb{Z}$: point-like defects (called hedgehogs) appear in nematics \cite{alexander2012colloquium}.
    \item $\pi_3(\mathcal{M})=\mathbb{Z}$: textures such as skyrmions, hopfions... appear in nematics \cite{chen2013generating,ackerman2014two}.
\end{itemize}
In the remainder of this work, we will narrow our purpose to line defects, mostly because of the specific role they play in different branches of physics. 

\section{Topological defects, a common pattern of nematics}\label{topo-part2}

\subsection{Typology of disclinations}
\begin{figure}
\begin{center}
\includegraphics[height=8.9cm]{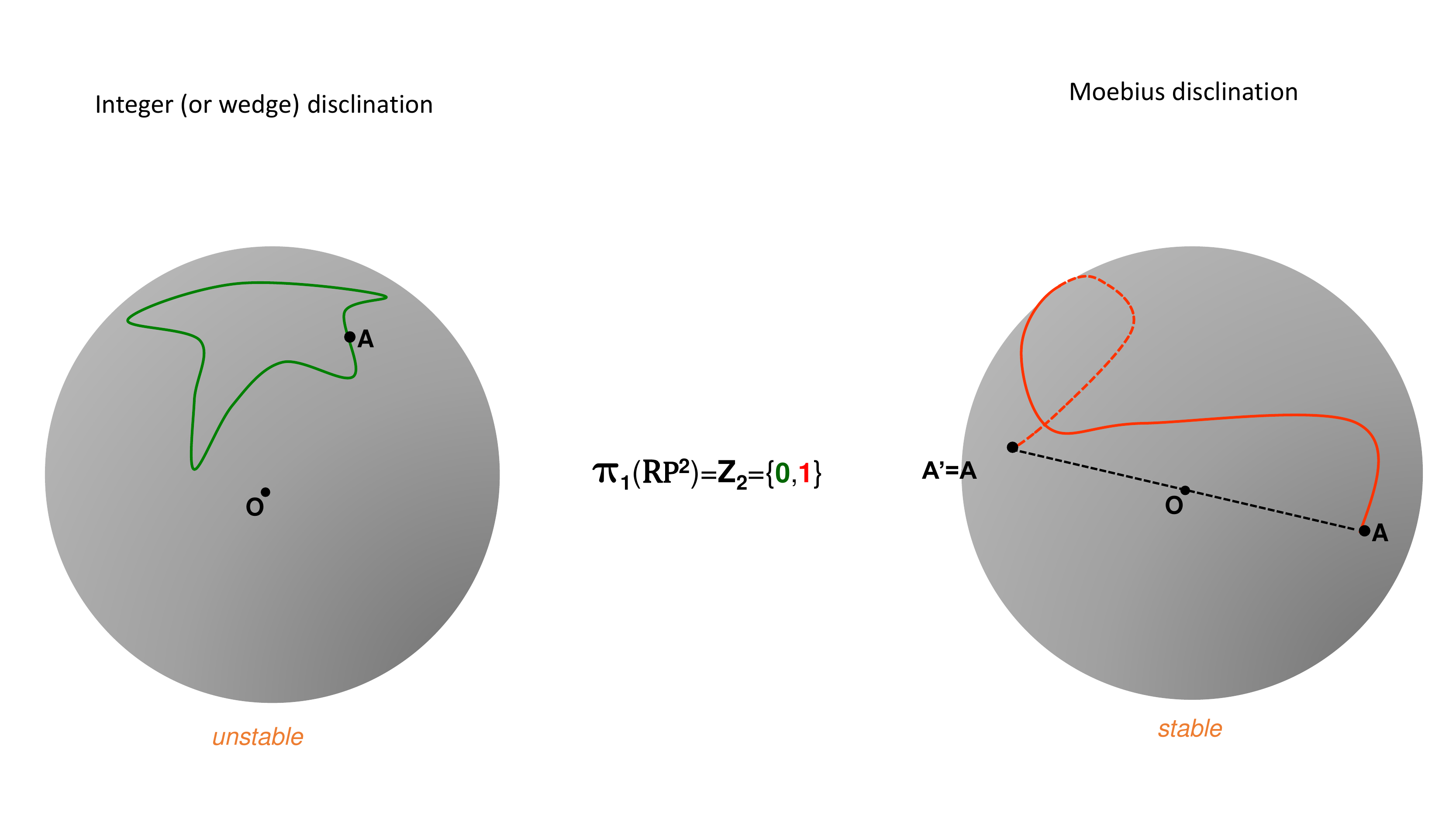} 
\caption{The two equivalence classes of $\pi_1$ on the Boy surface.}\label{Z2}
\end{center}
\end{figure}

Let us look closer at the homotopy content of the first fundamental group. The two equivalence classes for the addition law are $\pi_1(\mathcal{M})=Z_2=\{0,1\}$. The equivalence class associated to the neutral element $0$ corresponds to ordinary closed paths on the manifold (see the green loop on Fig \ref{Z2}): as they can be shrunk to a point, they correspond to removable singularities called wedge disclinations. But the particular structure of the Boy surface allows for another kind of loops: an open curve connecting two antipodal points (see the red curve on Fig \ref{Z2}). Obviously, there is no way to shrink the curve into a point while maintaining its closure: such singularity is topologically stable and are called M\oe bius disclinations. These two categories of defects combine according to the algebra of $Z_2$: $0+0=0$, $0+1=1$ and $1+1=0$. For instance in the latter case, this means that two open curves connected at the same antipodal points form a larger loop that can be shrunk to a point.      

To go further in the description of these singularities, let us build the Frank -- Oseen elastic free energy. Consider a given director field ${\vec n}_0(\mathbf{r})$. Deformations about that configuration are orthogonal to it as ${\vec n}_0.{\vec n}_0=1$, which leads to ${\vec n}_0.\delta{\vec n}=0$. To set things up, one considers ${\vec n}_0=\vec{e}_3$, which then demands $\delta{\vec n}=(\delta n_1,\delta n_2, 0)$. A Taylor expansion of $\delta{\vec n}$ reveals three deformation modes:
\begin{itemize}
    \item a splay mode in ${\vec\nabla}.{\vec n}$.
    \item a twist mode in ${\vec n}.({\vec\nabla}\times{\vec n})$
    \item a bend mode in ${\vec n}\times({\vec\nabla}\times{\vec n})$ 
\end{itemize}
Similarwise to the harmonic oscillator, the (simplest) Frank -- Oseen free energy density describing nematoelasticity writes as
\begin{equation}
    f\left[\vec n\right]=\frac{K_1}{2}\lvert{\vec\nabla}.\vec n\rvert^2+\frac{K_2}{2}\lvert\vec n.({\vec\nabla}\times\vec n)\rvert^2+\frac{K_3}{2}\lvert\vec n\times({\vec\nabla}\times\vec n)\rvert^2
\end{equation}
with $K_i$ the corresponding elastic constants. 

In the one-constant approximation, $K_1\approx K_2\approx K_3 =K= E_0/L$. Typically on has $E_0\approx 0.1$ eV, $L\approx 1$ nm, which gives $K\approx 10^{-11}$ N (verified for 5CB). Moreover, for planar configurations of the director field, ${\vec n}=\cos \psi\ \! \vec{e}_1+\sin \psi\ \! \vec{e}_2$. The Frank -- Oseen free energy density thus simplifies as 
\begin{equation}
    f\left[\vec n\right]=\frac{K}{2}\lvert{\vec\nabla}\psi\rvert^2 \label{FOfe}
\end{equation} 
Minimizing $f$ and retaining the director field configurations which do not depend on the radius finally gives $d^2\psi/d\theta^2=0$, that is:
\begin{equation}
    \psi(\theta)=m\theta+\psi_0
\end{equation}
where $m$ and $\psi_0$ are real constants. As the direction of ${\vec n}$ must be well-defined at each point, one has the constraint:
\begin{equation}
    \oint_{\theta}d\psi=2\pi m=k\pi\;\;\;k\in\mathbb{Z}
\end{equation}
as a result of $Z_2$ symmetry. This means that 1) $m$ is a winding number and 2) its values are restricted to $m=\pm 1/2, \pm 1, \pm 3/2, \pm 2...$. Substituting into (\ref{FOfe}), the free energy is now in $m^2$, which means that in practice, only distorsions with the lowest winding numbers are observed.

\begin{figure}
\begin{center}
\includegraphics[height=8.9cm]{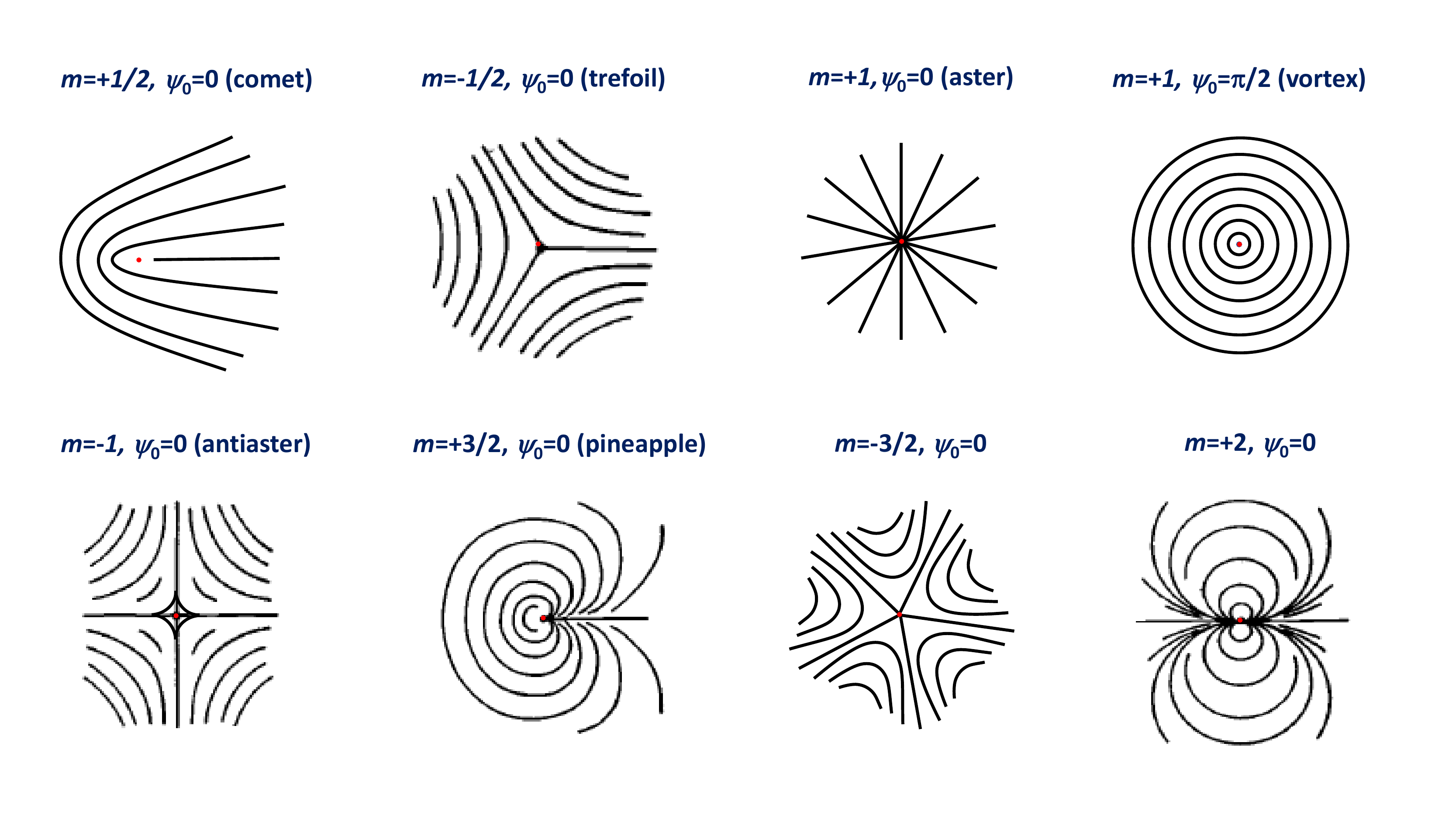} 
\caption{Examples of director field distributions for several planar line defects (the core singularity is in red). The antiaster and antivortex configurations are identical up to a $+\frac{\pi}{4}$ rotation. Adapted from \cite{oswald2005nematic}. }\label{usual-suspects}
\end{center}
\end{figure}

On Fig. \ref{usual-suspects}, several disclinations are plotted for different values of $m$ and $\psi_0$. The vortex and the aster configuration are toplogically equivalent, as they can be transformed into each other by a continuous rotation of the director field from $0$ to $\pi/2$: this means they belong to the same equivalence class. In fact, the director field of these defects can even deform continuously to relax into a non-singular configuration (escape into the third dimension): these defects are topologically unstable (this wouldn't be true for two-component order parameter $\vec n$~\cite{PhysRevE.72.031711}). As the same goes for $m=-1$ defects, one infers that disclinations of integer strengths belong to the trivial equivalence class (denoted as $N=0$) of $\pi_1$. On the contrary, disclinations of half-integer strengths are not topologically removable and they belong to the other equivalence class (denoted as $N=1$). Although $m$ is sometimes called the topological charge of the defect, it is the absolute value of $m$ that really matters for topology \footnote{The sign of $m$ can be distinguished by the rotation of the crossed polarisers. Indeed, polarising microscopy reveals Schlieren patterns, for which the number of dark brushes is $4\lvert m\rvert$0. Dark brushes from a positive (negative) defect rotate in the direction the same as (opposite to) that of the polarisers.}.

\subsection{Geometry of line defects}

From the standpoint of optics, nematics behave as uniaxial media. Their permittivity is $\varepsilon_{\parallel}$ in the direction of ${\vec n}$ and  $\varepsilon_{\bot}$ in  perpendicular planes. Seeking solutions in terms of plane waves show that such media support two optical modes: the ordinary wave, corresponding to a regular dispersion relation with refractive index $n_0=\sqrt{\varepsilon_{\bot}}$ and extraordinary wave, for which the ray index is anisotropic:
\begin{equation}
   N_e(\mathbf{r})=\sqrt{\varepsilon_{\bot}\cos^2\beta(\mathbf{r})+\varepsilon_{\parallel}\sin^2\beta(\mathbf{r})} 
\end{equation}
with $\beta(\mathbf{r})$ the angle between the director and the tangent vector to the ray. Propagation of the extraordinary light is ruled by Fermat -- Grandjean principle established in 1919 \cite{oswald2005nematic}: as ${\vec n}$ varies from point to point, so does $N_e$ and light is expected to follow curved paths. 

There is another well-known area in physics where light paths are curved: general relativity (for a modern textbook, see the excellent \cite{carroll2019spacetime}). Gravity is nothing more that the effect of spacetime curvature on the dynamics of massive and massless objects. For instance, in the vicinity of a massive star, light undergoes a gravitational lensing which bends its trajectory (this property was the first prediction of general relativity that was tested by Eddington in the same year 1919). The mathematical tool used to study curved spacetimes is differential geometry, most notably developed by Riemann. In the presence of curvature, all known laws of Euclidean geometry are shaken up. For instance, Pythagoras' theorem, usually written as $ds^2=dx^2+dy^2+dz^2$ now writes as \footnote{(For sake of simplicity, we omitted the time component, but in relativity, one must bear in mind time and space are put on an equal footing and the real interval must involve quadratic terms in $cdt$, with $c$ the speed of light.} 
\begin{equation}
    ds^2=g_{11}dx^2+g_{22}dy^2+g_{33}dz^2+2g_{12}dxdy+2g_{13}dxdz+2g_{23}dydz=\sum_{i,j=1,3} g_{ij}dx^idx^j
\end{equation}
The weighting coefficients $g_{ij}$ are the components of a second order tensor, called the metric, which can be seen as the multiplication table of the basis vectors chosen to describe the problem. In gravity, the metric is four-dimensional and is solution of Einstein's field equations that relate the geometry to the mass-energy content. 

How to extract a metric description for liquid crystals? 
This question was successfully addressed in \cite{satiro2006lensing,satiro2008deflection, pereira2013metric} and we will sum up here the recipe to be followed: 
\begin{enumerate}
    \item Consider a light path parametrized by $\ell$ and express the tangent vector in the Cartesian basis
    \begin{eqnarray}
    \vec{r}&=&r\cos\theta\ \!\vec{e}_x+r\sin\theta\ \!\vec{e}_y\\
        \vec{T}&=&\frac{d\vec{r}}{d \ell}=\left(\dot{r}\cos\theta-r\dot{\theta}\sin\theta\right)\vec{e}_x+\left(\dot{r}\sin\theta+r\dot{\theta}\cos\theta\right)\vec{e}_y
    \end{eqnarray}
    Here the dot is a shorthand notation for $d/d\ell$.
    \item Compute the components of ${\vec n}=\cos\psi\ \! \vec{e}_x+\sin\psi\ \!\vec{e}_y$ with respect to $\vec{T}$:
      \begin{eqnarray}
    \vec{T}.{\vec n}&=&\dot{r}\cos\left(\psi-\theta\right)+r\dot{\theta}\sin\left(\psi-\theta\right) \\
    \lvert\vec{T}\times{\vec n}\rvert&=&-\dot{r}\sin\left(\psi+\theta\right)+r\dot{\theta}\cos\left(\psi-\theta\right)
    \end{eqnarray}
    \item Replace in the Fermat -- Grandjean line element and see the defect metric appear (to spare lengthy calculations, one will consider the aster defect for which $\cos\beta=\dot{r}$ and $\sin\beta=r\dot{\theta}$): 
    \begin{eqnarray}
   ds^2=N_e^2(\vec{r})d\ell^2=\left(\varepsilon_{\bot}\dot{r}^2+\varepsilon_{\parallel}r^2\dot{\theta}^2\right) d\ell^2 =\varepsilon_{\bot}dr^2+\varepsilon_{\parallel}r^2 d\theta^2 
\end{eqnarray}
\end{enumerate}

\noindent Adding the $z$ term and performing a simple rescaling of the radial coordinate finally leads to the 3D line element
   \begin{eqnarray}
   ds^2=dr^2+\alpha^2r^2 d\theta^2+dz^2 \label{wedge-discli}
\end{eqnarray}
where $\alpha=\sqrt{\frac{\varepsilon_{\parallel}}{\varepsilon_{\bot}}}$. 

\begin{figure}[h!]
\begin{center}
\includegraphics[height=8.9cm]{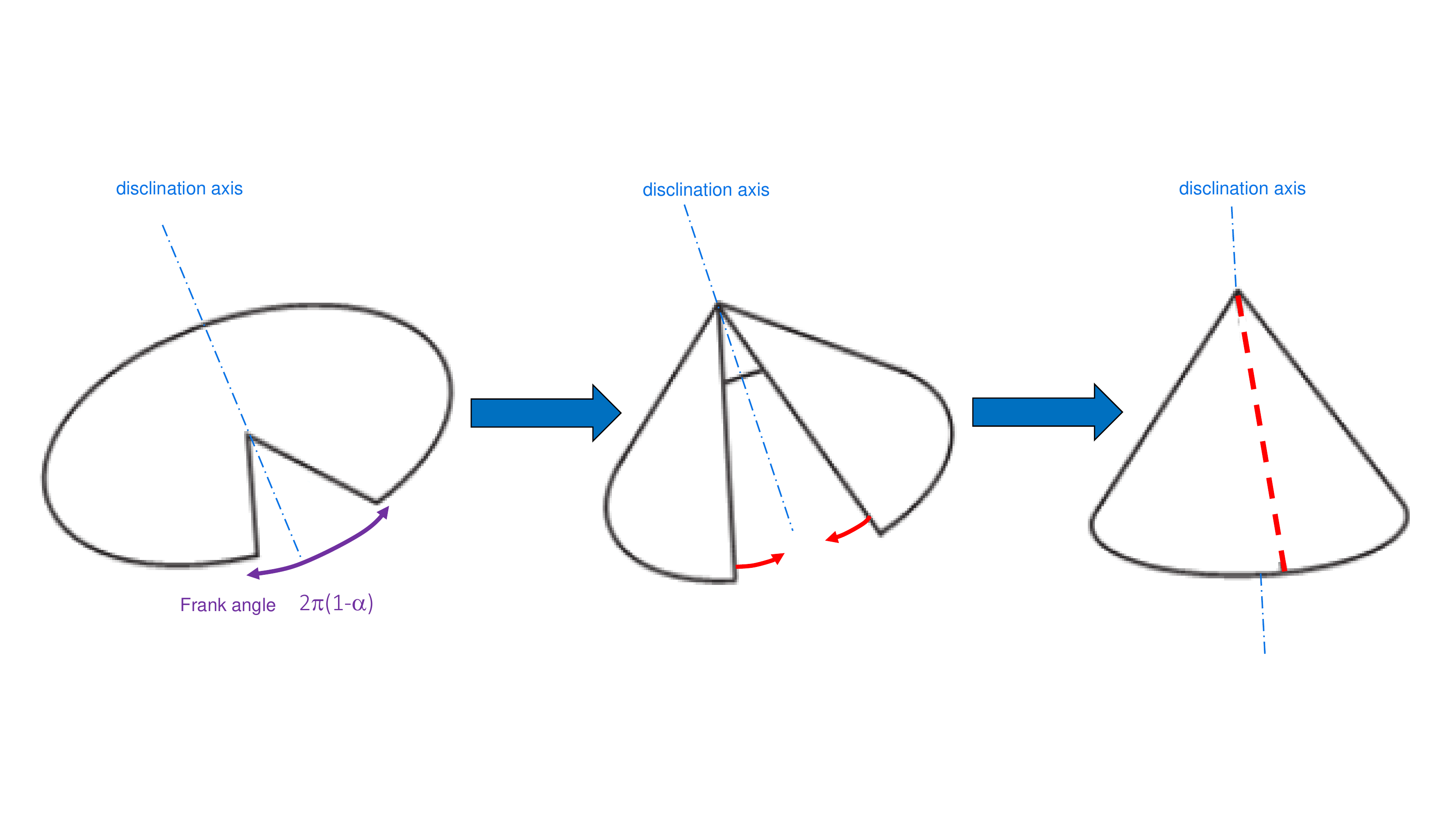} 
\caption{Volterra process for generating a wedge disclination with $\alpha<1$.}\label{volterra}
\end{center}
\end{figure}

What kind of geometry does this represent? A first hint is obtained by computing the Ricci scalar
\begin{equation}
  R(r)=\frac{1-\alpha}{\alpha r}\delta(r)  
\end{equation} 
The geometry is therefore flat everywhere but on the $z$-axis. Moreover, considering a circle of unit radius about the $z$-axis, the perimeter is given by $p=\oint ds=2\pi\alpha$. These two elements indicate that when $\alpha<1$ the geometry is conical, as described by a Volterra cut-and-glue process (see Fig. \ref{volterra}). 

For sake of completeness, we also report the line element corresponding to a general $(m,\psi_0)$-disclination line \cite{satiro2006lensing}:
\begin{eqnarray}
    ds^2&=&\left(\cos^2\left[(m-1)\theta+\psi_0\right]+\alpha^2\sin^2\left[(m-1)\theta+\psi_0\right]\right) \nonumber \\
    &&+\left(\sin^2\left[(m-1)\theta+\psi_0\right]+\alpha^2\cos^2\left[(m-1)\theta+\psi_0\right]\right) \nonumber \\ &&-\left(\alpha^2-1\right)\sin\left[2(m-1)\theta+2\psi_0\right]+dz^2
\end{eqnarray}

\subsection{Anholonomy}

Topological defects generate curved geometries and as such, one may expect many physical outcomes of incoming fields, the most natural being lensing effects \cite{satiro2006lensing, satiro2008deflection} and scattering \cite{pereira2013metric,fumeron2016retrieving}. This can be easily understood from the geodesic equations, which provide the shortest and autoparallel paths in a purely curved geometry (no torsion). Writing the parallel transport of the tangent vector along a curve gives
\begin{equation}
  \frac{d^2 x^i}{dt^2}+\Gamma^i_{jk} \frac{d x^j}{dt}\frac{d x^k}{dt}=0 
\end{equation}
where $\Gamma^i_{jk}$ are the Christoffel symbols of second kind:
\begin{equation}
  \Gamma^i_{jk}=\frac 12{g^{il}}\left(\partial_j g_{kl}+\partial_k g_{jl}-\partial_l g_{jk}\right)
\end{equation}
Geodesics correspond to the actual light paths and for the background metric (\ref{wedge-discli}), integration leads to
\begin{equation}
r(\theta)=\frac{C}{\alpha\sqrt{E}}\sqrt{1+\frac{\tan^2(\alpha\theta-F)}{2}}
\end{equation}
where $C, E$ and $F$ are integration constants (for graphical representations, see for instance \cite{de1998geodesics, satiro2006lensing}).

\begin{figure}
\begin{center}
\includegraphics[height=8.9cm]{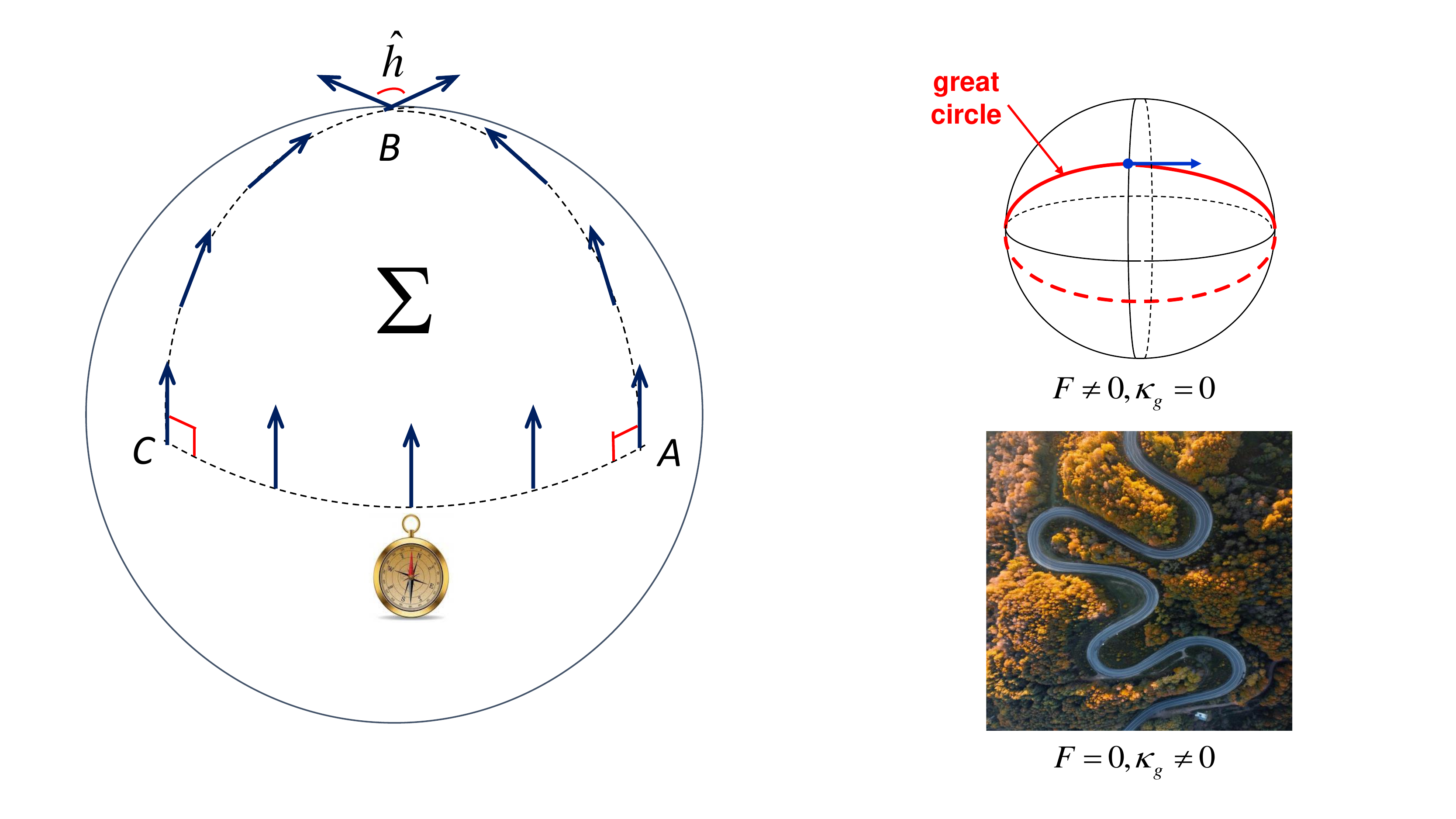} 
\caption{\textit{Left:} The world traveler problem. \textit{Right:} Difference between Gauss curvature (top) and geodesic curvature (bottom).}\label{globe}
\end{center}
\end{figure}

More subtle but yet related to parallel transport is the emergence of anholonomy effects, on which we will now focus our purpose. Let us consider first a practical problem (see Fig. \ref{globe}): a world traveler living in France (point A) wants to explore the north pole. He can either go directly to the north by following the prime meridian up to point C, or he can detour via Qu\'ebec (point B) and head north to C along the 70$^{\circ}$ W parallel. If he compares his compass needle in the two trips, he will discover a strange result: even if he parallel-transported in each circuit from the same starting and ending points, the direction vector at C is not the same. Stated otherwise: after a closed loop, a parallel-transported vector fails to recover its initial direction by a mismatch angle $\hat{h}$. This phenomenon is called anholonomy.

Intuitively, anholonomy can be understood as a topological effect. Girard's formula establishes the connection between the surface $\Sigma$ enclosed by circuit ABC and the Gauss curvature of the 2-sphere $F=1/R^2$:
\begin{equation}
    \Sigma=R^2\Omega=\frac{\hat{A}+\hat{B}+\hat{C}-\pi}{F}
\end{equation}
In the case depicted on Fig. \ref{globe}, this simplifies into $F\Sigma=\hat{C}$, which also turns out to be the mismatch angle $\hat{h}$: this latter is thus a measure of the Gaussian curvature \textit{F} of the surface $\Sigma$ bounded by the closed circuit. Such result is an basic outcome of the Ambrose -- Singer theorem: for a given connection on a vector bundle, the curvature corresponds to the surface density of holonomy. Now, Gaussian curvature is also related to the topology of the surface via Gauss -- Bonnet theorem:
\begin{equation}
    \iint_{\Sigma}FdS+\oint_{\partial \Sigma}\kappa_g ds=2\pi\chi
\end{equation}
Here, $\chi$ is the Euler -- Poincaré characteristic (topology) and $\kappa_g$ is the geodesic curvature. For the lost traveller problem, this simplifies into $\hat{h}=2\pi\chi$, which establishes the toppological origin of anholonomy.

A more formal and general expression of anholonomy has been obtained by Elie Cartan from the definition of parallel transport and is summed up in a modern form in \cite{carroll2019spacetime}. Let there be a path parametrized by $\lambda$ along which a vector $\bold{V}$ is parallel-transported. The parallel propagator $\Pi$ is defined as
\begin{equation}
    V^{\mu}(\lambda)=\Pi^{\mu}_{\;\rho}(\lambda) V^{\rho}(0)
\end{equation}
The parallel-transport condition applied to $\bold{V}$ writes as  
\begin{equation}
    \frac{d}{d\lambda}V^{\nu}(\lambda)=-\Gamma^{\nu}_{\sigma\mu}\frac{dx^{\sigma}}{d\lambda} V^{\mu}(\lambda)=A^{\nu}_{\;\mu}(\lambda)V^{\mu}(\lambda)
\end{equation}
where $\Gamma$ are Christoffel's connection symbols. This leads to the following equation for the propagator
\begin{equation}
    \frac{d}{d\lambda}\Pi^{\nu}_{\;\rho}(\lambda)=A^{\nu}_{\;\mu}(\lambda)\Pi^{\mu}_{\;\rho}(\lambda)
\end{equation}
This formally integrates into 
\begin{equation}
   \Pi^{\mu}_{\;\rho}(\lambda)=\delta_{\rho}^{\mu}+\int_0^{\lambda}A^{\mu}_{\;\sigma}(\eta)\Pi^{\sigma}_{\;\rho}(\eta)d\eta
\end{equation}
Similarly to what is done when establishing Dyson\'s formula, one can iterate the process. Instead of integrating over $n$-simplices, one integrates over $n$-cubes while keeping the product in the right order, to get the simpler expression: 
\begin{equation}
   \Pi^{\mu}_{\;\nu}(\lambda)=\text{P} \exp{\int_0^{\lambda}A^{\mu}_{\;\nu}}(\eta)d\eta
\end{equation}
where P is the ordering operator. On a loop $\gamma$ about a point $M$, the anholonomy writes explicitly as
\begin{equation}
   \Pi^{\mu}_{\;\nu}\left[\gamma\right]=\text{P} \exp\left(-\oint_{\gamma(M)} \Gamma^{\mu}_{\sigma\nu} dx^{\sigma}\right)
\end{equation}
As expected from Ambrose -- Singer theorem, to know the holonomy at every point of the manifold is equivalent to know the curvature at every point of the manifold: this property is heavily used in quantum loop gravity.  

Now, let us establish the anholonomy due to a disclination \cite{carvalho2007aharonov}. For a loop about the origin in a $z=C^{st}$ plane, only the polar connection symbol is retained:
\begin{equation}
    \Gamma_{\theta}=\frac{m}{\alpha}\left(\alpha^2\cos^2\left[(m-1)\theta+\psi_0\right]+\sin^2\left[(m-1)\theta+\psi_0\right]\right)
    \begin{pmatrix}
 0 & 1 \\
-1 & 0
  \end{pmatrix}
  \end{equation}
For instance, for the aster disclination, this reduces to $ \Gamma_{\theta}=\begin{pmatrix}
 0 & \alpha \\
-\alpha & 0
  \end{pmatrix}$  and the parallel propagator becomes:
\begin{equation}
     \Pi^{\mu}_{\;\nu}\left[\gamma\right]=\begin{pmatrix}
 \cos(2\pi\alpha) & -\sin(2\pi\alpha) \\
\sin(2\pi\alpha) & \cos(2\pi\alpha)
  \end{pmatrix}
  \end{equation}
When parallel-transporting a vector around it, the disclination causes an active rotation of angle $-2\pi\alpha$. All in all, the global mismatch angle when a vector describes a loop around the defect is $\hat{h}=2\pi-2\pi\alpha$. This means that disclinations generate a classical analog of the Aharonov -- Bohm effect. Indeed, the curvature is confined within the disclination line and vanishes everywhere else, but it has measurable effects affecting the phase of neighboring objects. This is one among many examples of what is generically called geometric or Berry phases, that is ``phases are not attributed to the forces applied onto the [quantum] system. Instead, they are associated with the connection of space itself''\cite{cohen2019geometric}. 

\section{Topological defects, a universal pattern of nature?}

\subsection{Biology}

\paragraph{A tentative definition} To begin with, we will discuss how a statistical physicist may describe biological systems. To do so, we will first need to address one question: what does it mean for a system to be at equilibrium? Figuring out what equilibrium means is not as simple as it appears at first glance. For instance, is the glass of my window at equilibrium or not? In his lectures on statistical physics, RP Feynman defined equilibrium as the situation when \guillemotleft \textit{all the fast things have happened but the slow things have not} \guillemotright \cite{richard1972feynman}. As handy as it sounds, this definition is yet incomplete as it blurs the demarcation between an equilibrium state and a non-equilibrium steady state. A better criterion to discriminate equilibrium from out-of-equilibrium situations is for a system to be crossed by fluxes of matter/energy/... exchanged with its surrounding. The glass of a window is nowadays understood as a metastable super-cooled liquid that do flow ought to gravity, but with relaxation times averaging $\tau\approx 10^9$ years (for water, $\tau$ is of the order of ms). The glass of my window is therefore crossed by fluxes of matter and it is in a non-equilibrium steady state. 

Living biological systems are out of equilibrium as well. For instance, a living cell consumes energy to maintain homeostasis (a non-equilibrium steady state) and perform mechanical works such as cellular division, membrane transport (water molecules sneak out the membrane as the phospholipid bilayer flexes and bends), motility (motion of flagella). From the standpoint of statistical mechanics, to be dead means to be at thermodynamic equilibrium. But contrary to the aforementioned non-living examples, the non-equilibrium steady state is not driven by external macroscopic fields, but by its internal components: animals in a flock, molecular motors and microtubules for cells... For the statistical physicist, biology is the study of active matter, e.g. non-equilibrium self-organized systems which do not couple trivially to the energy input  from their environment. Examples of active matter include colonies of bacteria, assemblies of myocyte cells, flocks of sheep, shoals of fish... Most of the time, the basic brick composing active matter are strongly anisotropic, such that the whole system displays a nematic order. Hence, at places where the average orientation of these bricks is ill-defined, topological defects arise. 

\paragraph{Active turbulence as a defect factory} Let us now investigate additional peculiarities of active matter by examining how it evolves in time. Usually, ordinary isotropic liquids obey the famous Navier -- Stokes equations \footnote{Finding general regular solutions of these equations is still one of the Millenium open problems listed by Clay Institute.}
\begin{equation}
    \rho\frac{D\bold{V}}{Dt}=\rho \bold{g}+{\boldsymbol\nabla}.\Bar{\Bar{\bm{\sigma}}}
\end{equation}
where the left-hand-side inertial terms include $D/Dt$, the material derivative, $\rho$ the mass density. The right-hand-side includes $\bold{g}$, the gravitational acceleration at Earth's surface, and $\Bar{\Bar{\bm{\sigma}}}=-p\Bar{\Bar {\mathbf I}}+2\eta \Bar{\Bar {\mathbf D}}$, the Cauchy stress tensor, which encompasses contributions due to the pressure $p$ and to the fluid viscosity $\eta$ ($\Bar{\Bar {\mathbf{D}}}$ stands for the strain rate tensor). The ratio between inertial terms to viscous forces defines a dimensionless quantity, the Reynold's number (originally introduced by George Stokes \cite{stokes1851effect}). At low Reynolds numbers, flows are laminar (regular and reversible) but for high Reynolds numbers, flows become turbulent (chaos and non-reversibility). 

In anisotropic liquids such as nematics, the fluid flow is furthermore coupled to the orientational order, which imparts them with unusual rheological features, such as the dewetting behavior of thin films, the presence of topological defects or backflow. Under the effect of an entering mass flow, the velocity field inside a nematic generates shear stresses that rotate the rod-like molecules and hence the director field. This mechanism is called advection. Conversely, let us submit a nematic at rest to an external electric field. For steric reasons, a rotation of the director exerts a shear stress on neighboring molecules which put them into motion: this is called backflow. The complete set of equations governing nematohydrodynamics (including advection and back-flow terms) form the Berris -- Edwards model\footnote{There is also another set forming the Ericksen -- Leslie equations, which are simpler but limited to uniaxial media and to smooth variations of the nematic ordering.}:
\begin{eqnarray}
    &&\rho\frac{D\bold{V}}{Dt}=\rho \bold{g}+{\boldsymbol\nabla}.\Bar{\Bar{\bm {\sigma}}} \label{BEP1} \\
    &&\frac{D\Bar{\Bar {\mathbf{Q}}}}{Dt}=\Gamma \Bar{\Bar {\mathbf {H}}}+\Bar{\Bar {\mathbf {S}}} +\lambda \Bar{\Bar{\mathbf {Q}}} \label{BEP2}
\end{eqnarray}
where $\Gamma$ is the rotational diffusion constant, $\Bar{\Bar{\mathbf {Q}}}$ is the Landau -- de Gennes order parameter tensor, $\Bar{\Bar{\mathbf {H}}}$ is the molecular field (driving relaxation towards a minimum of free energy) and $\Bar{\Bar{\mathbf {S}}}$ is the advection term. Compared to newtonian hydrodynamics, the Cauchy tensor in (\ref{BEP2}) includes an extra-term corresponding to backflow. Passive nematics can relax to configurations where the elastic energy is minimal: the medium then gets at rest after the decay of topological defects. 

A nematic becomes active as a result of many couplings with its environment (for a review see \cite{doostmohammadi2018active}). For instance, the complex motions of a shoal of sardines occur because each fish draws energy from marine plankton it ingested (energy couplings), but they can also respond to the presence of natural predators, local modifications of the sea properties... Every time, the energy fed at the individual scales (bacteria, cell, sheep, fish...) is transformed into organized motion at large scales. In many ways, this can be understood as a reverse Richardson cascade: contrary to ordinary turbulence where energy is transferred from large kinetic scales to small dissipative scales, the energy transfer occurs bottom-up. Active nematohydrodynamics is thus about the collective dynamics of energy-transducing anisotropic units, and the changes in Beris -- Edwards equations consist in adding two extra-terms: 1) a linear source term $\lambda\Bar{\Bar{\mathbf {Q}}}$ in (\ref{BEP1}) and 2) a new activity term $-\zeta\Bar{\Bar{\mathbf {Q}}}$ in the Cauchy tensor \cite{marenduzzo2007hydrodynamics}. These new terms result in a cyclic process: hydrodynamic instabilities generate surface singularities that fluctuate and decay into linear topological defects (mostly comets and trefoils), then defect-antidefect pairs annihilate (according to the algebra of $Z_2$), which in turn generates hydrodynamics instabilities etc. The proliferation of both line defects and vortices corresponds to a chaotic non-equilibrium steady state for low Reynolds numbers. We are next going to discuss several biological mechanisms in which disclinations play a crucial part.

\paragraph{Ethology, morphogenesis and oncology}

\begin{figure}
\begin{center}
\includegraphics[height=8.9cm]{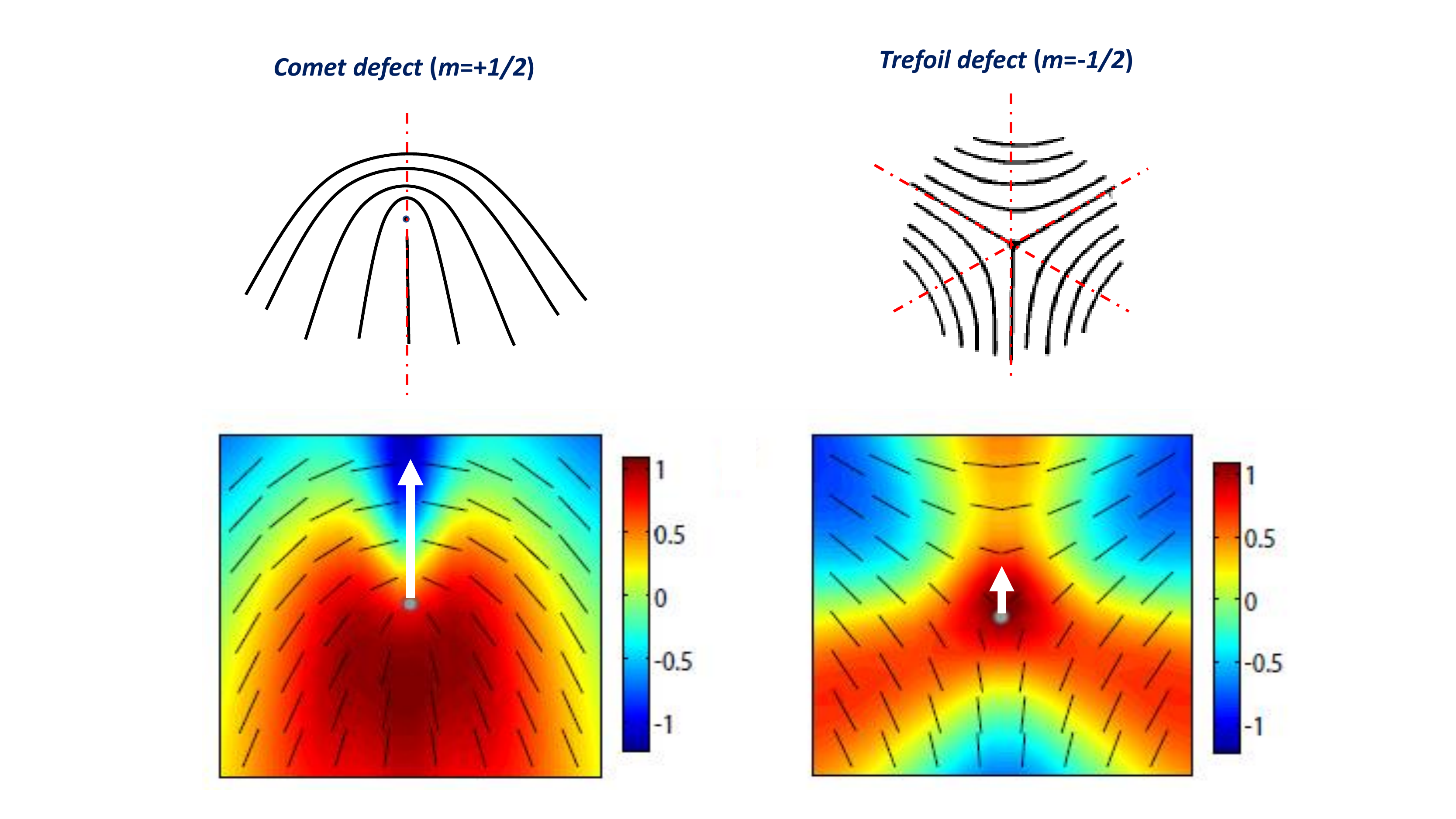} 
\caption{\textit{Upper-left:} Director field for a comet-like disclination (one symmetry axis in red dotted line). \textit{Lower-left}: Stress field around the comet (normalized units). Large net resulting force in white as a result of stress gradients (adapted from \cite{doostmohammadi2018active}). \textit{Upper-right:} Director field for a trefoil-like disclination (three symmetry axis in red dotted line). \textit{Lower-right}: Stress field around the trefoil (normalized units). Small net resulting force in white as a result of stress gradients (adapted from \cite{doostmohammadi2018active}).}\label{trefoil-comet}
\end{center}
\end{figure}

Comet- and trefoil-like defects behave quite differently in elastic media. Doostmohammadi et al. \cite{doostmohammadi2018active} both computed and measured the stress distribution in the vicinity of these two kinds of defects (see Fig \ref{trefoil-comet}). The directions where gradients are maximum correspond to the symmetry axis of the director field distributions. Averaging the stress contributions in the plane shows that there is a large net force associated to the comet which is oriented towards the head of the comet. On the contrary, for the trefoil, the resulting net force is very low (theoretically, it vanishes if the sample is axisymmetric). In the last decades, our understanding of the delicate mechanisms involved in the functioning of organs has made substantial progress. Cells are now understood as highly-sensitive mechanotransductive units, displaying an orientational order ought to the elongated structures (actin and intermediate filaments, microtubules) forming the cytoskeleton. It must be remarked that actin and microtubule may also form polar organic materials, in which the formation of $\pm 1$ wedge disclinations is energetically favored. Hence, a topological defect causing a singularity in the orientational order, i.e. in the stress distribution, is likely to trigger several biological responses, driving processes such as the dynamics of animal groups, morphogenesis or disease initiation \cite{doostmohammadi2021physics}.

Colonies of bacteria, insect swarms and schoals of fish all bahave as active nematics, as they are essentially assemblies of anisotropic motile units. The effect of M\oe bius defects on bacteria populations seems well-captured. Refs.~\cite{peng2016command,genkin2017topological} found that in a lyotropic liquid crystal (disodium cromoglycate),  the populations of \textit{Bacillus subtilis} -- a common plant growth-promoting rhizobacteria from the soil --  use to swim from trefoil-like defect cores to comet-like defect core. Similarly, \cite{copenhagen2021topological} showed that colonies of \textit{Myxococcus xanthus}, another common rod-like bacteria living in the soil, may form new cell layers at +1/2 defects and holes at -1/2 defects (competition between different colonies also showed that +1/2 defects help a colony prevailing over another \cite{meacock2021bacteria}). For larger animals such as insects, fish and birds, the role attributed to defects is not as clear because of non-linear processes \cite{schaller2013topological}. A recent work led with a colloidal liquid suggests that the remarkable flow stability of flocking matter could come from the self-advection and density gradient around $-1$ topological defects \cite{chardac2021topology}.

Morphogenesis and vesicle growth have recently been understood as processes driven by topological defects \cite{fardin2021living,guillamat2022integer}. The reason for this is at the interplay between mechanics, soft matter physics and topology. Many living systems have an outer layer rich in actin filaments and microtubules, which provides them with an orientational order. Ought to the Poincar\'e -- Brouwer theorem (sometimes called the "hairy ball theorem"), lines covering any closed curved surface present at least one singular point, where tangent vectors are ill-defined (the global topological charge in a spheroid is set at +2): this explains the presence of cyclonic vortices at the Earth's surface or rebellious cowlicks on someone's head. For a biological organism topologically equivalent to a 2-sphere, the hairy ball theorem states that topological defects are unavoidable on its surface. Keber et al. \cite{keber2014topology} showed that defect sites on a spherical vesicle rich in microtubules were precursors for the growth of protrusions. Cell differentiation and swirling protrusions have recently been reported to be driven by integer defects (spirals and asters) in myoblast monolayers \cite{blanch2021integer,guillamat2022integer}, known to be rich in actin and myosin. \cite{maroudas2021topological} investigated the biophysics of hydra, a small freshwater predatory animal that is almost immortal as it can regenerate each organ (head, foot, mouth, tentacles). After being cut into pieces, each part folds into a spheroid that supports topological defects. Ought to the nematic actin organization, differentiation and regeneration occur at protrusions growing from defect sites, the topological charge being associated to one and only kind of organ (+1 for mouth, foot and tentacle, $-1/2$ for the base).

M\oe bius defects may also play a important part in oncology. The transcription coactivator called YAP (Yes-Associated Protein) is known to promote tumorigenesis, metastasis and even chemotherapy resistance \cite{camargo2007yap1,lamar2012hippo,wang2013mutual,marti2015yap,warren2018yap,shen2018hippo}. YAP is also critical in cell death mechanisms (such as apoptosis and ferroptosis) and cell extrusion from biological tissues (see \cite{cheng2022biology} for a recent review). Hence, in principle, it should be possible to use YAP  to trigger cancer cell death. And this is precisely where topological defects come into play: high levels of compressive stresses at the head of comet-like defects are known to translocate YAP from the nucleus to the cytoplasm and to trigger cellular death and extrusion \cite{saw2017topological,doostmohammadi2021physics}. However, the journey to defect-engineered therapies based on YAP deactivation promises to be long, as competing mechanisms involving defect-induced cell motions have recently been reported to impede malignant cell clearance \cite{zhang2021topological}.

\subsection{Cosmology and cosmic strings}

\paragraph{Phase transitions in cosmology} According to the Standard Hot Big Bang Model, about 13.8 billion-years ago, our Universe was in an extremely hot dense state, consisting in a quark-gluon plasma. In the framework of grand unified theory, the four fundamental interactions were supposed to be unified, the corresponding "superforce" being invariant under the element of a grand unified gauge symmetry group \textit{G}. 
Then, the universe expansion played the role of a gigantic Joule -- Thomson expansion, which caused large temperature drops likely to give rise to cosmological phase transitions with spontaneous gauge symmetry breaking (SSB). After the last of these transitions occurred (electroweak phase transition) at about $10^2$ GeV, the electroweak force split up into the electromagnetic force and the weak nuclear force, which corresponds to the gauge symmetry $SU(3)_c\times U(1)_{em}$. As in condensed matter, the topology associated to the symmetry-breaking pattern provides informations on the possible topological defects that may appear. Only the terminology is changing: instead of the order parameter space, one speaks of the vacuum manifold, which is the set of Higgs field configurations minimizing energy modulo gauge transformations. Jeannerot et al determined the homotopy content corresponding to all eligible groups \textit{G} likely to decay below $10^{16}$GeV into $SU(3)_c\times SU(2)_L \times U(1)_Y$. Their conclusion leaves no doubt concerning the formation of cosmic strings: \guillemotleft \textit{among the SSB schemes which are compatible with high energy physics and cosmology, we did not find any without strings after inflation} \cite{jeannerot2003generic} \guillemotright.

\begin{figure}
\begin{center}
\includegraphics[height=8.9cm]{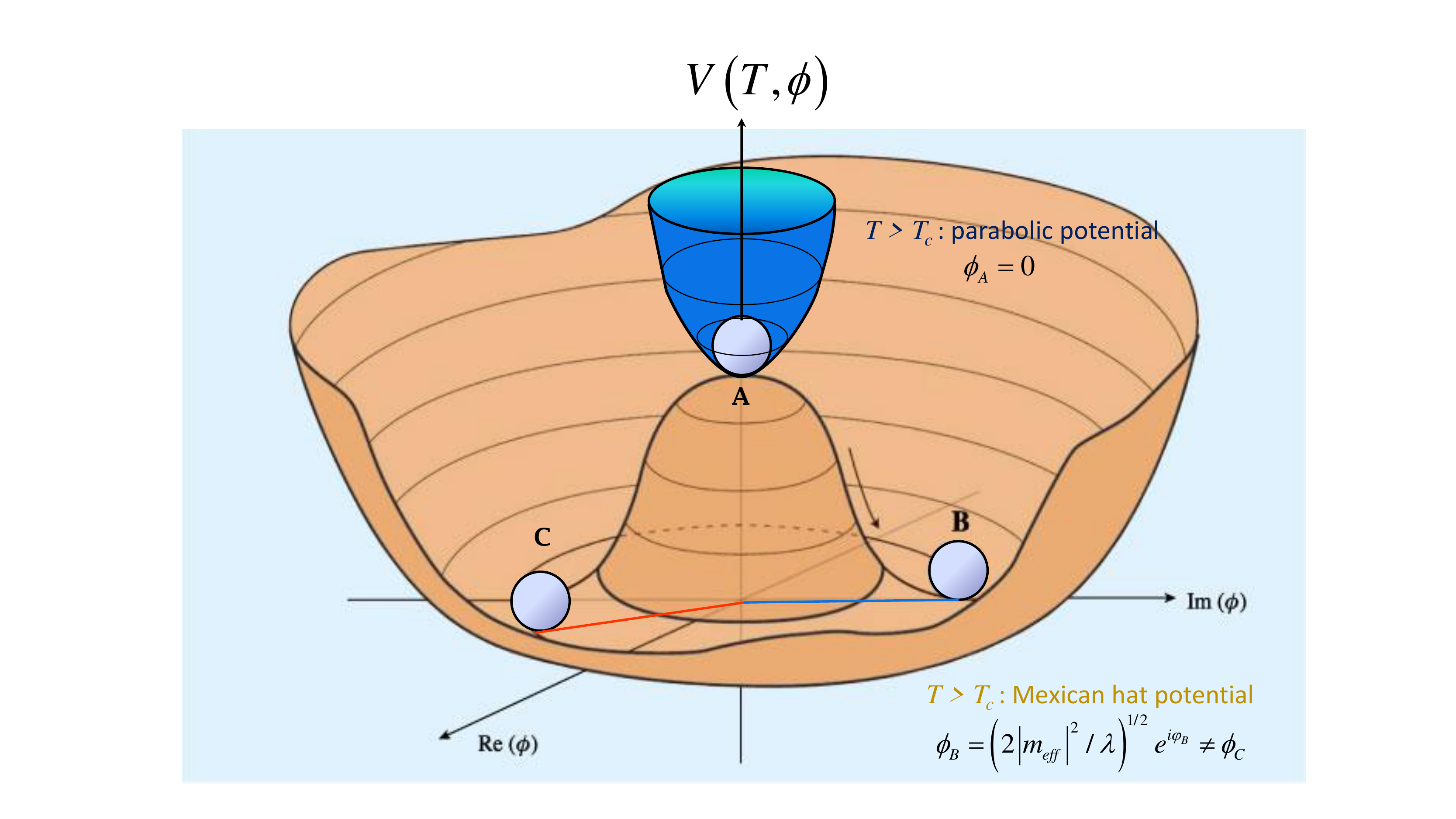} 
\caption{A toy-model from thermal field theory: the abelian Higgs model. At high temperatures, the parabolic potential gives a vanishing Higgs field $\phi$ and the gauge symmetry is $U(1)$. At low temperatures, the parabolic potential gives a vanishing Higgs field $\phi$ and the gauge symmetry is broken into $I$. Adapted from \cite{ellis2016historical}}.\label{Higgs-toy-model}
\end{center}
\end{figure}

Of prime importance is the phase of cosmic inflation that presumably happened at the very beginning of the universe \cite{guth1981inflationary}. It consists in an extremely rapid expansion (typically a factor $10^{26}$ within $10^{-32}$ seconds) likely to solve cosmological riddles such the horizon and flatness problems~\cite{Mukhanov:2005sc}. From the point of view of statistical physics, inflation is nothing more than a tremendous quench and it is likely to promote the formation of topological defects. Kibble \cite{kibble1976topology} and later Zurek \cite{zurek1996cosmological} proposed a mechanism now known as the Kibble -- Zurek scenario (KZS) describing the different steps of this quench. It starts with a nucleation process, similarly to what happens at the isotropic-nematic phase transition, but the orientational order comes from the phase choice of a complex scalar field generically called a Higgs field. The fast temperature drop due to inflation causes the Higgs field to locally take a non-vanishing vacuum expectation value and hence to make a phase choice (see Fig. \ref{Higgs-toy-model}). This leads to an ordered region called a protodomain (analog to a nematic spherulite). Then the protodomains grow in size and eventually coalesce, but as they were not causaly connected, the choices for the Higgs phases do not necessarily match. Singularities of the Higgs phase appear  where the boundaries of protodomains finally meet, giving rise to line-like singularities called cosmic strings.

\paragraph{Statistical physics of a cosmic string network} Kibble and Zurek remarked that these phase transitions involves two competing velocity scales: 1) the velocity $v_f$  at which the field fluctuations propagate and 2) the velocity $v_p$ at which the parameter ruling the phase transition (here the temperature) varies. When $v_p > v_f$, the system is quenched and the parameters describing the resulting distribution of cosmic defects depend on the quench time $\tau_q$ in the form of scaling laws. For instance, the correlation length, the relaxation time, the average density of defects and the variance of the net winding number $\sigma$ are respectively given by:   
\begin{eqnarray}
\xi(t)\sim\left|\frac{t-t_c}{\tau_q}\right|^{-\nu}\;\;\;\;\;\;\;\;\;\;\tau(t)\sim\left|\frac{t-t_c}{\tau_q}\right|^{-\mu}\;\;\;\;\;\;\;\;\rho(t)\sim\left(\frac{t}{\tau_q}\right)^{-\alpha}\;\;\;\;\;\;\;\;\;\sigma\sim N^{1/4} 
\end{eqnarray}
where $t_c$ is the time when the transition occurs \cite{peter2009primordial} and $N$ is the total number of defects in the region under investigation. 
\medskip 

In the 1990s, several works \cite{chuang1991cosmology,bowick1994cosmological,digal1999observing,kibble2007phase,mukai2007defect,repnik2013symmetry} showed that the KZS, originally developped for cosmology, was also accurately describing disclinations in nematics with the very same scaling coefficients. For instance, this model predicts that the density of strings scales as $\rho \sim \left(t/\tau_q\right)^{\alpha}$ with a critical exponent $\alpha_{th}=0.5$, and measurements done by \cite{chuang1991cosmology} with 5CB indeed gave $\alpha_{th}=0.51\pm0.04$. The exponent characterizing the correlation between defects and antidefects is expected to be 1/4 and was measured at $0.26\pm 0.11$ \cite{digal1999observing}. From the standpoint of statistical physics, phase transitions in cosmology and in liquid crystals seem to belong to the same universality class. But the family resemblance goes further. Networks of cosmic strings and networks of disclinations also share the same intersection processes: 1) when two line defects intertwine, they can reconnect the other way as they cross (intercommutation) \cite{vilenkin1985cosmic,chuang1991cosmology} and 2) when one line defect self-intersects, it generates a loop \cite{brandenberger1994topological,duclos2020topological}.

\paragraph{Spacetime near a Nambu -- Goto string} The analogy between disclinations and cosmic strings goes even deeper. The simplest cosmic defects one may expect in the universe are called Nambu -- Goto strings: they consist in delta-distributed concentrations of mass-energy and they can be pictured as infinitely straight and thin objects (the thickness of a realistic cosmic string is estimated at $10^{-28}$ cm). As required by thermal field theory and general relativity, the geometry around a Nambu -- Goto string (in units where $c=1$) is described by the Vilenkin's line element \cite{vilenkin1985cosmic}: 
\begin{equation}
    ds^2=-dt^2+dr^2+\left(1-4 G\mu\right)^2r^2d\theta^2+dz^2 \label{Vilenkin}
\end{equation}
where $\mu$ is the string mass-energy density estimated at about 10 million billion tons per meter! The space part of this element is identical to (\ref{wedge-discli}) and it corresponds to a conical geometry with a removed Frank angle  (typically, for a grand unified scale string, this angle is a few seconds of arc). As curvature is confined to the string axis, spacetime is locally ﬂat away from the string, which exerts no gravitational pulling onto neighboring objects. Such object may of course generates a Berry phase in a similar fashion to their soft-matter cousin \cite{ford1981gravitational}.

From the standpoint of the soft-matter physicist, Nambu -- Goto strings can be understood as the cosmic counterparts of wedge disclinations. How to make sense of such incredible similarity? For the most part, this question is still open, but a noticeable attempt to address it was done in \cite{simoes2010liquid}. In essence, the reason is that equations of nematoelasticity have the form as the spatial sector of Einstein’s field equations, with the elastic-stress tensor playing the role of the energy momentum tensor. 

As the analogy between gravity and nematoelasticity does not concern time components, the dynamics of a cosmic defect cannot be directly mapped with those of a disclination. The motion of disclinations is classical (typically a few $\mu$m per second) and friction-dominated, whereas cosmic strings are ultra-relativistic and dissipation mechanisms are due to radiation of gravitational waves. As noticed in \cite{chuang1991cosmology}, this is intimately connected to the nature of the broken symmetries: in particle physics, these latter are gauged (or internal) whereas in liquid crystals, broken symmetries are geometrical \footnote{Still, this difference is also the reason why we have used two different notations for vectors, $\bf r$ for the ordinary space vectors and $\vec n$ for the order parameter degrees of freedom.}.

\subsection{Miscellanea}

As a final bouquet, we will now discuss other fields in science where the key role played defects has been identified. Let us start with the wonder material of solid state physics: graphene. This material was identified in 2004 by Geim and Novoselov, and eversince its amazing chemical and physical  (electrical, mechanical, thermal...) properties propelled it to the scientific forefront. In fact, the perfect flatness of graphene plays a crucial role in its unusual behavior as local curvature  modifies the local density of electronic states \cite{cortijo2007effects}. As graphene supports whole-integer disclination dipoles (pentagon-heptagon configurations known as Stone -- Wales defects), one may expect to taylor the physical properties of graphene from distributions of topological defects: this is the emerging field of defect engineering.  Tayloring graphene curvature with defects may even be relevant for medical sciences, as the curvature dependence of biomolecular adsorption may help discriminating between different molecules and removing harmful molecules in disease treatments \cite{ni2019topological}. Recently, tayloring transport in low dimesional systems has been refined to the scale of quasi-particles and immobile gauge modes known as fractons were shown to be intimately connected to disclinations \cite{pretko2018fracton}. Defect engineering is also considered in soft matter ought to its high flexibility \cite{fumeron2014principles,barros2018concurrent}.

Line defects also provides the answer to some fundamental questions in geology: as Earth's mantle is a solid layer made of rocks, how can it slip and move? Indeed, plasticity usually comes from the mobility of screw dislocations, but they are not sufficient to account for the rheology of the mantle. The main constituent of the upper mantle is olivine, at about 60-70\%. In \cite{cordier2014disclinations}, it was shown that this layer was extremely rich in disclinations. Any applied shear on olivine-rich rocks induces grain-boundary migration mediated by disclination motions, explaining the motions of the Earth's mantle. This was confirmed by high-resolution electron backscattering diffraction.

Last but not least suprising is the central situation of line defects in high energy physics. In \cite{deser1984three}, Deser, Jackiw and 't Hooft proposed a model in 2+1 dimension where particles interacting gravitationally are described by a gas of conical defects, the mismatch angle being proportional to the particle's mass. This model was later extended to 3+1 dimensions by 't Hooft \cite{t2008locally} but this time, matter particles are represented by a gas of piecewise straight string segments. An important feature of this model is that particles can have positive or negative masses, i.e. $\alpha<1$ and $\alpha>1$ Frank angles. This suggests that nematics could be used to perform high energy physics experiments, in the spirit of the analogue gravity game plan \cite{barcelo2011analogue,jacquet2020next}. The last word will be on quark physics. Topological insulators are currently raising a considerable attention because of their unusual electric properties. The main difference between ordinary insulators and topological insulators (such as Chern, Kane -- Mele...) is the band inversion: spin-orbit couplings link valence and conduction bands. Ought to this connectivity, the surface states are topologically protected from local perturbations. Topological insulators may support disclinations in the bulk, which disrupt the lattice structure. This is responsible for the emergence of (quasi)particles of fractional charge, trapped at the defect location \cite{ortix2021electrons,peterson2021trapped}. This is likely to be of interest to simulate the elusive quarks, which have charge $-1/3e$ or $2/3e$.   

\section{Conclusion}

Topological defects consist in regions of ordered systems where the order locally broke down (for instance after a phase transition), and as such they are among the most widespread structures in science. They carry generic topological and geometrical properties likely to impact a wide variety of physical processes, therefore enabling mutual cross-fertilisation between different domains. For instance, the Kibble mechanism, initially designed for cosmology, was firstly tested with 5CB, a standard liquid crystal. Conversely, the existence of M\oe bius defects in nematics suggests the search for more general cosmic defects in our universe (such as Alice strings). 

Future prospects on topological defects look promising, both on  the fundamental front and on the technological front. On the fundamental front, the search for cosmic strings is still in wait of a first robust observational evidence: so far, the NANOGrav collaboration only report stochastic gravitational-wave background signals compatible with, among others, networks of cosmic strings with a string parameter $G\mu$ in the range $\left[ 10^{-10.0}, 10^{-10.7}\right]$ \cite{blanco2021comparison}. Investigations on the role of defects in biology is still running at full tilt and a recent review on open problems can be found in \cite{ardavseva2022topological,doi:10.1021/acsnano.1c01347}.  One of the most challenging questions may probably be to understand how the mechanical stresses due to defects couple to biochemical signalling.

Another booming field of research is the emerging field of defect engineering. On the one hand, defects induce a change in the background geometry experienced by low-energy excitations. On the other hand, soft-matter systems, in particular liquid crystals, are indeed well-renowned contenders to build functionalized devices. Ought to advances in surface treatments, this has come to the point where it is now possible to prepare well-organised assemblies of topological defects from photopatterning \cite{guo2021photopatterned} or micro-well structures \cite{sakanoue2022controlled} (see \cite{jangizehi2020defects} for a review). Therefore, curvature/torsion can be taken advantage of to taylor the outgoing fluxes for applications in mechanical engineering (acoustics, electronics, thermotronics, optics..., see e.g. \cite{refId0,Fumeron_2017EPL,doi:10.1063/1.4921310,unpublished_review}). Such perspectives are also under investigations in biology, where the possibility to control the dynamics of defects pairs was shown for tissues prepatterned from photoaligned liquid crystal elastomer \cite{turiv2020topology}. Whatever the field of applications, it leaves no doubt that taming topological defects is only at its first steps.


\end{document}